\documentclass[11pt]{amsart}

\usepackage{amscd}
\usepackage{amsmath}
\usepackage{graphicx}
\usepackage{amsfonts}
\usepackage{amssymb}
\begin{document}

\title[Constructing separable  states]{Constructing separable  states in infinite-dimensional systems by operator matrices}

\author{Jinchuan Hou and Jinfei Chai}

\address{ ~Department of Mathematics, Taiyuan University of Technology, Taiyuan 030024, China}
\email[JHou]{houjinchuan@tyut.edu.cn; jinchuanhou@aliyun.com}
\email[JChai]{449287913@qq.com}

\begin{abstract}

We introduce a class of  states so-called semi-SSPPT (semi super
strong positive partial transposition) states  in
infinite-dimensional bipartite systems by the Cholesky decomposition
in terms of operator matrices  and show that every semi-SSPPT state
is separable. This gives a method of constructing separable states
and   generalizes the corresponding results in [Phys. Rev. A
\textbf{77}, 022113(2008); J. Phys. A: Math. Theor. 45 505303
(2012)]. This criterion is specially convenient to be applied when
one of the subsystem is a qubit system.

\end{abstract}

\maketitle

\section{Introduction and main result}

Entanglement is  an important resource  in quantum information
processing and quantum computation~\cite{Nielsen}. However, the
detection of entanglement is one of the most difficult task in this
area and much more effort had been paid to on this research field
~\cite{Horodecki,Guhne,Hou2,Stormer,Hou,Augusiak,Namiki,Hou3,Hou4,Hou6,Zhaomingjing}.
Recall that,  mathematically, a quantum state $\rho$ (positive
operator of trace 1) acting on a separable complex Hilbert space
$H=H_A\otimes H_B$  is called \emph{separable} if it can be written
as the form
\begin{eqnarray}
\rho=\sum_ip_i\rho_i^A\otimes\rho_i^B,\quad \sum_ip_i=1,~
p_i\geq0 \label{d}
\end{eqnarray}
or can be approximated in the trace-norm by the states of the above
form, where $\rho_i^A$ and $\rho_i^B$ are respectively quantum
states of subsystem A and B \cite{Werner,HP2}. Otherwise, $\rho$ is
called inseparable or \emph{entangled}. A separable state with the
form as in Eq.(\ref{d}) is called \emph{countably separable}
\cite{Holevo,Guo3}. If $\dim H_A\otimes H_B<+\infty$, it is known
that all separable states $\rho$ acting on $H_A\otimes H_B$ are
countably separable \cite{HP2}. But, in the infinite-dimensional
case, there exists separable states which are not countably
separable \cite{Holevo}.

One of the most famous and convenient criteria for detecting
entanglement is the
 positive partial transpose (PPT) criterion proposed by
Peres and Horodecki \cite{Peres,Horodecki2} which asserts that if a
quantum state $\rho$ acting on   $H_A\otimes H_B$ is separable, then
its partial transposes are positive operators, that is,
$\rho^{T_{A/B}}\geq0$. There are entangled PPT states except for
those in $2\otimes 2$ and $2\otimes 3$ systems \cite{Horodecki2}.
So, it is important to know which PPT states are separable.

In this paper we present a method to construct a class of PPT states
on infinite-dimensional bipartite systems by an operator-matrix
trick and show that such states are separable.

In a bipartite system A+B described by $H_A\otimes H_B$ with $\dim
H_A\otimes H_B=+\infty$, let $\{|i_a\rangle\}$ and $\{|j_b\rangle\}$
be any orthonormal bases of $H_A$ and $H_B$, respectively. Denote by
$E^a_{kl}=|k_a\rangle\langle l_a|$ and $E^b_{kl}=|k_b\rangle\langle
l_b|$. Then any state $\rho$ acting on $H_A\otimes H_B$ can be
represented by
\begin{eqnarray}
\rho=\sum_{k,l}^{\dim H_A}E^a_{kl}\otimes B_{kl}=\sum_{k,l}^{\dim
H_B}A_{kl}\otimes E^b_{kl},\label{a}
\end{eqnarray}
where $B_{kl}$s are trace class operators on $H_B$ and the series
converges in trace-norm \cite{Guo}; that is,
\begin{eqnarray}
\rho=\left(\begin{array} {c|c|c|c|c}
B_{11}&B_{12}&B_{13}&\cdots&\cdots \\ \hline
B_{21}&B_{22}&B_{23}&\cdots&\cdots \\ \hline
\vdots&\vdots&\ddots&\vdots&\cdots\\ \hline
B_{m1}&B_{m2}&B_{m3}&\cdots&\cdots\\ \hline
\vdots&\vdots&\vdots&\cdots&\ddots
\end{array}\right) = \left(\begin{array} {c|c|c|c|c}
A_{11}&A_{12}&A_{13}&\cdots&\cdots \\ \hline
A_{21}&A_{22}&A_{23}&\cdots&\cdots \\ \hline
\vdots&\vdots&\ddots&\vdots&\cdots\\ \hline
A_{n1}&A_{n2}&A_{n3}&\cdots&\cdots\\ \hline
\vdots&\vdots&\vdots&\cdots&\ddots
\end{array}\right)\label{w}
\end{eqnarray}
under the given bases. Take operator sequences $\{X_i\}_{i=1}^{\dim
H_A}$ and $\{S_{ij}\ :\ 1\leq i<j\leq {\dim H_A}\}$ on $H_B$ so that
the operator matrix  (infinite if $\dim H_A=\infty$)  of the form
\begin{eqnarray}
X=\left(\begin{array}{c|c|c|c|c|c}
X_1&S_{12}X_1&S_{13}X_1&\cdots&S_{1m}X_1&\cdots \\ \hline
0&X_2&S_{23}X_2&\cdots&S_{2m}X_2&\cdots \\ \hline
\vdots&\vdots&\ddots&\vdots&\vdots&\cdots\\ \hline
0&0&0&X_{m-1}&S_{m-1,m}X_{m-1}&\cdots\\ \hline 0&0&0&0&X_m&\cdots\\
\hline \vdots&\vdots&\vdots&\vdots&\vdots&\ddots
\end{array}\right)
\end{eqnarray}
is a Hilbert-Schmidt operator on $H_A\otimes H_B$, that is,
$$
{\rm Tr}(X^\dag X)=\sum_{i} {\rm Tr}(X_i^\dag X_i)+\sum_{i<j}{\rm
Tr}(X_i^\dag S_{ij}^\dag S_{ij}X_i)<\infty,
$$
then
\begin{eqnarray}
\rho^b_X=\frac{1}{{\rm Tr}(X^\dag X)}X^\dag X
\end{eqnarray}
is a bipartite state in $H_A\otimes H_B$. One can construct
$\rho^a_X$ in the same way. For convenience, we call
$\{X_i\}_{i=1}^{\dim H_A}$ and $\{S_{ij}\ :\ 1\leq i<j\leq {\dim
H_A}\}$ Cholesky operators, and say Eq.(5) is a Cholesky
decomposition $\rho_X^a$. These terminologies come from the fact
that every block matrix has a Cholesky decomposition of the form in
Eq.(5)

{\bf Definition 1.} {\it A state $\rho\in{\mathcal S}(H_A\otimes
H_B)$ is called a semi-SSPPT state up to part B if it has a Cholesky
decomposition as in Eq.(5) and the associated Cholesky operators
$\{S_{ij}\ :\ 1\leq i<j\leq {\dim H_A}\}$ satisfying the condition
\begin{eqnarray}
[S_{ki},S_{kj}^\dag]=0,\quad k<i\leq j. \label{u}
\end{eqnarray}
The semi-SSPPT states up to part A are defined similarly. A state is
called a semi-SSPPT state if it is a semi-SSPPT states up to part B
or a semi-SSPPT states up to part A.}

It is easily checked that every semi-SSPPT state is PPT. The
following is our may result.

{\bf Theorem 1.}\quad {\it Let $\rho\in\mathcal{S}(H_A\otimes H_B)$
be a state with $\dim H_A\otimes H_B\leq\infty$. If $\rho$ is
semi-SSPPT, then $\rho$ is separable.}

The terminology SSPPT (super strong positive partial transpose)
comes from \cite{Chruscinski} for finite-dimensional systems and
\cite{GH} for infinite-dimensional systems, where the additional
assumption ``{\it every $S_{ij}$ is diagonalizable}" is required.
The main result in \cite{Chruscinski, GH} shows that the SSPPT
states are countably separable. However, though the condition Eq.(6)
ensures that every $S_{ij}$ is a normal operator, we know that there
are many normal operators on infinite-dimensional Hilbert spaces
that are not diagonalizable. Thus the above theorem 1 generalizes
the result in \cite{Chruscinski, GH} greatly.

The proof of theorem 1 will be presented in Appendix B. We point
out, our proof of theorem 1 needs new mathematical tools including
introducing a concept of SOT-separability for bounded positive
operators and establishing a Radon-Nikodym type theorem for spectral
measure, which we present in Appendis A.

\section{Corollaries and examples}

Theorem 1 provides an easier way of constructing separable states in
infinite-dimensional bipartite systems.

{\bf Example 1.} Assume $\dim H_A=n\leq\infty$ and $\dim
H_B=m\leq\infty$. Let $\{X_i\}_{i=1}^n$ and $\{S_i\}_{i=1}^n$ be two
sequences of operators on $H_B$ such that $\sum_{i=1}^n{\rm
Tr}(X_i^\dag X_i)=\frac{1}{2}$, $S_i$ normal and ${\rm Tr}(X_i^\dag
S_i^\dag S_iX_i)\sum_{i<j}^n\frac{1}{2^{2j}}={\rm Tr}(X_i^\dag X_i)$
for each $i\in\{1,2,\ldots,n\}$. Let $S_{ij}=\frac{1}{2^j}S_i$ for
$j>i$ and let $X$ be the operator matrix as in Eq.(4). Then,
$${\rm Tr}(X^\dag X)=\sum_{i=1}^n{\rm
Tr}(X_i^\dag X_i)+\sum_{i=1}^n {\rm Tr}(X_i^\dag S_i^\dag
S_iX_i)\sum_{i<j}^n\frac{1}{2^{2j}}=\frac{1}{2}+\frac{1}{2}=1.
$$
Thus, $\rho=X^\dag X$ is a   state in $H_A\otimes H_B$ and is
separable by theorem 1. In the case $\dim H_A=n=\infty$, as
$\sum_{i<j}^\infty\frac{1}{2^{2j}}=\frac{1}{3}\frac{1}{2^{2i}}$, one
may choose normal operator $S_i$ so that ${\rm Tr}(X_i^\dag S_i^\dag
S_iX_i)=\frac{9}{2}$ for each $i$ to ensure that $\sum_{i=1}^\infty
{\rm Tr}(X_i^\dag S_i^\dag
S_iX_i)\sum_{i<j}^n\frac{1}{2^{2j}}=\frac{1}{2}$.

The following are some corollaries of theorem 1 which generalize the
corresponding results in \cite{GH} from finite-dimensional systems
to infinite-dimensional systems, and also illustrates the use of
theorem 1 to detect the separability of a state in the case when
$\min\{\dim H_A,\dim H_B\}=2$. Note that every trace-class operator
acting on an infinite-dimensional Hilbert space can not be
invertible; so the corollary in \cite{GH} is not applicable to
infinite-dimensional case.

Assume that  $\dim H_A=2$ (or $\dim H_B=2$) and
$\rho\in\mathcal{S}(H_A\otimes H_B)$. Then $\rho$ can be written in
\begin{eqnarray}
\rho=\left(\begin{array}{cc}
\rho_{11}&\rho_{12}\\
\rho_{21}&\rho_{22}\end{array}\right)\quad {\rm (or }\ {\rho}=
\left(\begin{array}{cc}\tilde{\rho}_{11}&\tilde{\rho}_{12}\\
\tilde{\rho}_{21}&\tilde{\rho}_{22}\end{array}\right))
\end{eqnarray}
up to part B (or, up to part A).

 {\bf Corollary 1.}\quad {\it Let $\rho$ be a state as in Eq.(7). If
there is a  Cholesky decomposition $\rho=X^\dag X$ up to part B/A
with $ X=\left(\begin{array}{cc}
X_{1}&S_{12}X_1\\
0&X_2\end{array}\right)$ such that $X_1$ has dense range and
$\rho^{T_{A/B}}=Y^\dag Y$ with $Y= \left(\begin{array}{cc}
X_{1}&S_{12}^\dag X_1\\
0&X_2\end{array}\right)$, then $\rho$ is separable.}

\noindent{\bf Proof.}\quad   Since $(Y^\dag
Y)^{T_{A/B}}=(\rho^{T_{A/B}})^{T_{A/B}}=\rho=X^\dag X$, one gets
$X_1^\dag S_{12}S_{12}^\dag X_1= X_1^\dag S_{12}^\dag S_{12}X_1$,
which entails that $S_{12}S_{12}^\dag= S_{12}^\dag S_{12}$ as the
range of $X_1$ is dense. So, $\rho$ is semi-SSPPT and thus, by
Theorem 1, is separable.\hfill$\square$

{\bf Corollary 2.}\quad {\it Let $\rho$ be a state as in Eq.(7).
 Then any one of the following conditions implies that $\rho$ is separable.}

 (1)  $\rho_{11}\geq\rho_{22}$ (or $\tilde{\rho}_{11}\geq\tilde{\rho}_{22}$).

 (2)  $\rho_{22}\geq\rho_{11}$ (or $\tilde{\rho}_{22}\geq\tilde{\rho}_{11}$).

{\bf Proof.} Assume that $\rho$ satisfies the condition
$\rho_{11}\geq\rho_{22}$. In this case,  $\dim H_A=2$ and $
\rho=\left(\begin{array}{cc}
\rho_{11}&\rho_{12}\\
\rho_{21}&\rho_{22}\end{array}\right)$ with
$\rho_{ij}\in\mathcal{T}(H_B)$ and $\rho _{22}\leq \rho_{11}$. We
shall show that $\rho$ is semi-SSPPT and hence is separable by
Theorem 1.

Since $\rho\geq 0$ and $\rho_{22}\leq \rho_{11}$, there are
contractive operators $T$, $S$ on $H_B$ with $\ker T\cap\ker
S\cap\ker S^\dag\supseteq \ker \rho_{11}$ such that
$\rho_{12}=\sqrt{\rho_{11}}T\sqrt{\rho_{22}}$ and
$\sqrt{\rho_{22}}=\sqrt{\rho_{11}}S=S^\dag
\sqrt{\rho_{11}}$~\cite[Theorem 1.1]{Hougao}, here, $\ker L$ denotes
the null space of the operator $L$.

 Let
$S_{12}=TS^\dag$ . Then we have
$\rho_{12}=\sqrt{\rho_{11}}S_{12}\sqrt{\rho_{11}}$. Note that
\begin{eqnarray*}
\sqrt{\rho_{11}}S_{12}^\dag
S_{12}\sqrt{\rho_{11}}=\sqrt{\rho_{22}}T^\dag T\sqrt{\rho_{22}}\leq
\rho_{22}.
\end{eqnarray*}
Let
\begin{eqnarray*}
X_2=[\rho_{22}-\sqrt{\rho_{11}}S_{12}^\dag
S_{12}\sqrt{\rho_{11}}]^{\frac{1}{2}},
\end{eqnarray*}
\begin{eqnarray*}
X=\left(\begin{array}{cc} \sqrt{\rho_{11}} &S_{12}\sqrt{\rho_{11}}\\
0& X_2\end{array}\right) \quad{\rm and}\quad
Y=\left(\begin{array}{cc} \sqrt{\rho_{11}}
&S_{12}^\dag\sqrt{\rho_{11}}\\ 0& X_2\end{array}\right).
\end{eqnarray*}
Then
\begin{eqnarray*}
\rho=X^\dag X=\left(\begin{array}{cc} {\rho_{11}}
&\sqrt{\rho_{11}}S_{12}\sqrt{\rho_{11}}\\
\sqrt{\rho_{11}}S_{12}^\dag\sqrt{\rho_{11}}&
\sqrt{\rho_{11}}S_{12}^\dag S_{12}\sqrt{\rho_{11}}+X_2^\dag
X_2\end{array}\right)
\end{eqnarray*}
and
\begin{eqnarray*}
\rho ^{T_A}=&Y^\dag Y=\left(\begin{array}{cc} {\rho_{11}}
&\sqrt{\rho_{11}}S_{12}^\dag\sqrt{\rho_{11}}\\
\sqrt{\rho_{11}}S_{12}\sqrt{\rho_{11}}& \sqrt{\rho_{11}}S_{12}
S_{12}^\dag\sqrt{\rho_{11}}+X_2^\dag X_2\end{array}\right).
\end{eqnarray*}
Since $[\rho^{T_A}]^{T_A}=\rho$, we get
\begin{eqnarray*}
\sqrt{\rho_{11}}S_{12}
S_{12}^\dag\sqrt{\rho_{11}}=\sqrt{\rho_{11}}S_{12}^\dag
S_{12}\sqrt{\rho_{11}}.
\end{eqnarray*}
On the other hand, $\ker T\cap\ker S\cap\ker S^\dag\supseteq \ker
\rho_{11}$ ensures that $\ker S_{12}\supseteq \ker
\sqrt{\rho_{11}}$. This entails that $S_{12} S_{12}^\dag=S_{12}^\dag
S_{12}$, that is, $\rho$ is semi-SSPPT.

Other cases con be dealt with similarly.
\hfill$\square$\\

We give an example to illustrate how to apply Corollary 2.\\

{\bf Example 2.}\quad Assume $\dim H_A=2$ and $\dim H_B\leq\infty$.
For any operators $\rho_{11}, D$ and $T$ acting on $H_B$ with
$\rho_{11}$ a positive trace-class operator, $\|D\|\leq 1$ and
$\|T\|\leq 1$. Obviously, by the corollary 2 the state $\rho\in
S(H_A\otimes H_B)$ constructed by
$$\rho=\frac{1}{{\rm
Tr}(\rho_{11}+\sqrt{\rho_{11}}DD^\dag
\sqrt{\rho_{11}})}\left(\begin{array}{cc}\rho_{11}&
\sqrt{\rho_{11}}T[\sqrt{\rho_{11}}DD^\dag
\sqrt{\rho_{11}}]^{\frac{1}{2}}\\
{[\sqrt{\rho_{11}}DD^\dag\sqrt{\rho_{11}}]^{\frac{1}{2}}T^\dag\sqrt{\rho_{11}}}&
\sqrt{\rho_{11}}DD^\dag \sqrt{\rho_{11}} \end{array}\right)$$ is
separable since
\begin{eqnarray*}
\rho_{11}\geq\sqrt{\rho_{11}}DD^\dag \sqrt{\rho_{11}}=\rho_{22}.
\end{eqnarray*}

\section{Conclusions}

 In terms of the Cholesky decomposition and local commutativity, we introduce a notion
 of semi strongly super positive partial
transpose (semi-SSPPT) states and establish a criterion of
separability: if a quantum state in an infinite-dimensional
bipartite system  is semi-SSPPT, then it is  separable. This
criterion generalizes the corresponding results in
\cite{Chruscinski, GH} and gives a way of constructing separable
states by operator matrices. This criterion is specially convenient
to be applied when one of the subsystem is a qubit system.  To prove
this criterion, we establish a Radon-Nikodym type theorem for the
spectral measure. However, our Radon-Nikodym type theorem is stated
in term of unbounded operators and the strong operator topology
(SOT) convergence. This forces us to introduce a notion of
SOT-separability for positive operators acting on  tensor product of
two Hilbert spaces,
 and show that a state
is separable if and only if it is SOT-separable.  These results
together enable us to give a proof of the main criterion. Our
discussion also reveals that introduce and study separability for
positive operators acting on tensor product of Hilbert spaces are
helpful for solving some problems raised in quantum information
theory.

{\bf Acknowledgments} {This work is partially supported by Natural
Science Foundation of China (11671294).}

\section*{{\bf Appendix A}: \\ SOT-separability for positive operators and
Radon-Nikodym type theorem for spectral measure}

To prove  theorem 1, we need generalize the concept of separability
from states to  bounded positive operators and Radon-Nikodym theorem
from vector mesasures in Hilbert space to the spectral measures.

Denote by $ {\mathcal B}(H)$ and $ {\mathcal B}_+(H)$  the set of
all   bounded linear operators and the set of all positive bounded
operators acting on a complex Hilbert space $H$, respectively.

{\bf Definition A.1.} A positive operator $T\in{\mathcal
B}_+(H_A\otimes H_B)$ is called SOT-separable if there exist
positive operators $\{A_k\}\subset{\mathcal B}_+(H_A)$ and
$\{B_k\}\subset{\mathcal B}_+(H_B)$ such that
$$
T=\sum_k A_k\otimes B_k \eqno(A.1)
$$
or if $T$ is the limit of the operators of the form as in Eq.(8)
under the strong operator topology (briefly, SOT). Otherwise, $T$ is
said to be SOT-inseparable.

We remark that, if the sum in Eq.(8) is a series, we mean that the
series is convergent under SOT. It is obvious that $0$ is
SOT-separable and, if $\dim H_A\otimes H_B<\infty$, then   a nonzero
positive operator $T$ is SOT-separable if and only if $\frac{1}{{\rm
Tr}(T)}T$ is a separable quantum state.

Denote by ${\mathcal S}_{\rm SOT}(H_A\otimes H_B)$ the set of all
SOT-separable operators, which is a SOT-closed convex cone in
${\mathcal B}(H_A\otimes H_B)$.

The following is a SOT-separability criterion, which is similar to
the entanglement witness criterion for states.

{\bf Proposition A.1.} {\it A  positive operator $ T$ is
SOT-inseparable if and only if there is a self-adjoint operator
$W\in{\mathcal B}(H_A\otimes H_B) $ of finite rank such that}

(1) {\it ${\rm Tr}(W(A\otimes B))\geq 0$ holds for any
$A\in{\mathcal B}_+(H_A)$ and $B\in{\mathcal B}_+(H_B)$;}

(2) ${\rm Tr}(WT)< 0$.

{\bf Proof.}  Since  ${\mathcal S}_{\rm SOT}(H_A\otimes H_B)$ is
 a SOT-closed convex subset of ${\mathcal B}(H_A\otimes H_B)$,
by Hahn-Banach theorem, $T\not\in {\mathcal S}_{\rm SOT}(H_A\otimes
H_B)$ if and only if there exists a SOT-continuous linear functional
$\phi$ on ${\mathcal B}(H_A\otimes H_B)$ and a real number $c$ such
that ${\rm Re}(\phi(S))\geq c$ for all $S\in {\mathcal S}_{\rm
SOT}(H_A\otimes H_B)$ but ${\rm Re}(\phi(WT))<c$. As $\phi$ is
SOT-continuous, there are vectors $x_1, \ldots, x_r; y_1,\ldots,
y_r\in H_A\otimes H_B$ with $r<\infty$ such that
$\phi(X)=\sum_{i=1}^r \langle y_i|X|x_i\rangle $ holds for all $X\in
{\mathcal B}(H_A\otimes H_B)$. Let $E=\sum _{i=1}^r
|x_i\rangle\langle y_i|$. Then $E\in {\mathcal B}(H_A\otimes H_B)$
is a finite-rank operator which satisfies $ \phi(X)={\rm Tr}(EX) $
for all $X$ (Ref., for example, \cite{Co}). If $X$ is self-adjoint,
that is, if $X^\dag=X$, then $ \phi(X)^*={\rm Tr}(EX)^*= {\rm
Tr}((EX)^\dag)={\rm Tr}(E^\dag X)$ and hence ${\rm Re}(\phi(X))={\rm
Tr}(WX)$, where $W={\rm Re}(E)=\frac{1}{2} (E+E^\dag)$. It follows
that ${\rm Tr}(WS)={\rm Re}(\phi(S))\geq c$ for all $S\in {\mathcal
S}_{\rm SOT}(H_A\otimes H_B)$ and ${\rm Tr}(WT)={\rm
Re}(\phi(T))<c$. Note that $0\in{\mathcal S}_{\rm SOT}(H_A\otimes
H_B)$. So we must have $c\leq 0$. We assert that ${\rm Tr}(WS)\geq
0$ holds for any $S\in{\mathcal S}_{\rm SOT}(H_A\otimes H_B)$. If,
on the contrary, there is $S\in{\mathcal S}_{\rm SOT}(H_A\otimes
H_B)$ so that ${\rm Tr}(WS)=a<0$. As ${\mathcal S}_{\rm
SOT}(H_A\otimes H_B)$ is a convex cone, $tS\in{\mathcal S}_{\rm
SOT}(H_A\otimes H_B)$ for any $t>0$. Thus we have $ta={\rm
Tr}(W(tS))\geq c$ for all $t>0$, which is a contradiction since
$ta\to -\infty$ when $t\to \infty$. Therefore, we have found a
self-adjoint operator $W$ of finite rank such that (1) and (2) hold.

Conversely, if there is some self-adjoint operator $W$ of finite
rank such that (1) and (2) hold, then ${\rm Tr}(WS)\geq 0$ holds for
all $S\in{\mathcal S}_{\rm SOT}(H_A\otimes H_B)$ since $\phi :
X\mapsto {\rm Tr}(WX)$ is a SOT-continuous linear functional. Thus
$\phi$ separates strictly $T$ and ${\mathcal S}_{\rm SOT}(H_A\otimes
H_B)$. So $T\not\in{\mathcal S}_{\rm SOT}(H_A\otimes H_B)$.
\hfill$\Box$

Next we discuss the relationship between separability and
SOT-separability for a quantum state. Notice that, generally
speaking, though a sequence $\{T_n\}$ of positive trace-class
operators converges to a state $\rho$ under SOT, one can not assert
that $\{T_n\} $ converges to $\rho$ under the trace-norm. For
instance, let $\{a_n\}_{n=1}^\infty$ be a sequence of positive
numbers so that $\sum_{n=1}^\infty a_n =a<\infty$. For any $n$, let
$T_n={\rm diag} (t_1,t_2,\ldots, t_k\ldots)$ with $t_k=0$ if $k\leq
n$ and $t_{n+m}=a_m$. Then $\{T_n\}_{n=1}^\infty$ is a sequence of
positive trace-class operators and $T_n\to 0$ in SOT. However, ${\rm
Tr}(T_n)=a$ which does not converge to $0$. In addition, for a state
$\rho$, let $T^\prime_n=\rho+T_n$. Then $T_n^\prime\to  \rho$ under
SOT but ${\rm Tr}(T^\prime_n)=1+a$ does not converge to ${\rm
Tr}(\rho)=1$. Hence, $\{T^\prime_n\}$ does not converge to $\rho$ in
trace-norm. This suggests that a SOT-separable state may not be
separable. However, the following result reveals  surprisedly   that
this is not the case.

{\bf Proposition A.2.} {\it Let $\rho\in {\mathcal S}(H_A\otimes
H_B)$ be a state. Then $\rho$ is separable if and only if $\rho$ is
SOT-separable.}

{\bf Proof.} Clearly, $\rho$ is separable implies that $\rho$ is
SOT-separable by the definitions as the convergence in trace-class
norm implies the convergence in SOT.

Conversely, assume that $\rho$ is inseparable (i.e., entangled). We
have to show that $\rho$ is also SOT-inseparable. Let
$\{|i_a\rangle\}$ and $\{|j_b\rangle\}$ be arbitrarily  given
orthonormal bases for $H_A$ and $H_B$, respectively. Let $P_k$ and
$Q_k$ be finite-rank projections on $H_A^{(k)}$ and $H_B^{(k)}$, the
span of $\{|i_a\rangle\}_{i=1}^k$ and the span of
$\{|j_b\rangle\}_{j=1}^k$, respectively. If $k\geq \dim H_A$ (or
$k\geq \dim H_B$), let $P_k=I_A$ (or $Q_k=I_B$) with $I_A$ the
identity operator on $H_A$. Let
$$\rho_k=\frac{1}{{\rm Tr}((P_k\otimes Q_k)\rho(P_k\otimes
Q_k))}(P_k\otimes Q_k)\rho(P_k\otimes Q_k).$$ Obviously,
$\rho=\|\cdot\|_{\rm Tr}$-$\lim _{k\to\infty} \rho_k$. As $\rho$ is
inseparable, there exists infinitely many $k$ so that $\rho_k$ is
inseparable (otherwise, $\rho$ should be separable). Take a such
$k$. Then $\rho_k$ can be regarded as an inseparable state in the
finite-dimensional system $H_A^{(k)}\otimes H_B^{(k)}$. Thus, by the
entanglement witness criterion, there is a self-adjoint operator
$W_k$ on $H_A^{(k)}\otimes H_B^{(k)}$ such that ${\rm
Tr}(W_k\rho_k)<0$. Let $W=(P_k\otimes Q_k)W_k(P_k\otimes Q_k)$. $W$
is a self-adjoint operator of rank $\leq k^2<\infty$ and is an
entanglement witness for $\rho$ because for any pure states
$P_A\in{\mathcal S}(H_A)$ and $Q_B\in{\mathcal S}(H_B)$, $${\rm
Tr}(W(P_A\otimes Q_B))={\rm Tr}(W_k((P_kP_AP_k)\otimes
(Q_kQ_BQ_k)))\geq 0$$ and $${\rm Tr}(W\rho)={\rm Tr}((P_k\otimes
Q_k)W_k(P_k\otimes Q_k)\rho)={\rm Tr}(W_k\rho_k)<0.$$

For any $A\otimes B\in{\mathcal B}_+(H_A\otimes H_B)$, $(P_k\otimes
Q_k)(A\otimes B)(P_k\otimes Q_k)=(P_kAP_k)\otimes (Q_kBQ_k)$ is
either zero or a positive multiple of a finite rank separable state.
So, we still have
$${\rm Tr}(W(A\otimes B))\geq 0.$$ This implies by Proposition A.1
that $\rho$ is SOT-inseparable, as desired. \hfill$\Box$.

Now, let us turn to the question of establishing Radon-Nikodym type
theorem for spectral measure. Let   $\Omega$ be a nonempty set,
$\mathcal B$ be a $\sigma$-algebra of subsets of $\Omega$, $H$ be a
Hilbert space.  Recall that a spectral measure for $(\Omega,
{\mathcal B}, H)$ is an operator-valued function  $E: {\mathcal
B}\to {\mathcal B}(H)$ such that

(i) for each $\Delta$ in $\mathcal B$, $E(\Delta)$ is a projection;

(ii) $E(\emptyset)=0$ and $E(\Omega)=I$;

(iii) for $\Delta_1,\Delta_2\in{\mathcal B}$,
$E(\Delta_1\cap\Delta_2)=E(\Delta_1)E(\Delta_2)$.

(iv) if  $\{\Delta_i\}\subset {\mathcal B}$ are pairwise disjoint
sets, then $E(\cup_i \Delta_i)=\sum_i E(\Delta_i)$, here the sum
converges in SOT (Ref. \cite{Co}).

{\bf Proposition A.3.} (The Radon-Nikodym type theorem for spectral
measure) {\it Let $H$ be a complex  Hilbert space,  $\Omega$ be a
nonempty set, $\mathcal B$ be a $\sigma$-algebra of subsets of
$\Omega$. Assume that $E$ is a spectral measure for
$(\Omega,\mathcal B, H)$ and $\mu$ is a positive measure on
$(\Omega, {\mathcal B})$. If $E\ll \mu$, that is, if
$\mu(\Delta)=0\Rightarrow E(\Delta)=0$, then there is an
operator-valued function $D: \Omega \to {\mathcal B}(H)$ such that
$\langle x|D(\omega)|x\rangle\geq 0$ a.e. $\mu$ and
$$E(\Delta)x={\rm( B)}\int _\Delta D(\omega)x{\rm d} \mu_\omega
$$
holds for every $x\in H$ and $\Delta\in{\mathcal B}$, where ${\rm(
B)}\int _\Delta$ means the Bochner integral. }

We remark that  $D(\omega)$ may take a unbounded operator.

{\bf Proof.} Recall that a Banach space $X$ is said to have the
Radon-Nikodym Property (RNP) if for any finite positive measure
space $(\Omega, {\mathcal F}, \mu)$ and vector-valued measure
$F:{\mathcal F}\to X$, if $F\ll \mu$, then there exists a Bochner
integrable vector-valued function $f:\Omega\to X$ such that
$F(\Delta)={\rm (B)}\int_\Delta f(\omega){\rm d} \mu_\omega$ holds
for any $\Delta\in{\mathcal F}$. Not every Banach space has RNP.
However it is well-known that every Hilbert space has RNP (ref.
\cite{Cl,Ar}).

Now let $(\Omega, {\mathcal B}, \mu)$ be a finite positive  measure
space and $(\Omega, {\mathcal B}, H,E)$ be a spectral measure space
so that $E\ll \mu$. We remark here that we can not use the result in
\cite{Ar} because the spectral measure is not $\sigma$-bounded. For
any vector $x\in H$, it is clear that $F_x: {\mathcal B}\to H$
defined by $F_x(\Delta)=E(\Delta)x$ is a $H$-valued measure
satisfying $F_x\ll \mu$ and $\sigma$-boundedness. As $H$ has RNP,
there is a Bochner integrable vector-valued function $D_x:\Omega\to
H$ such that $E(\Delta)x=F_x(\Delta)={\rm (B)}\int_\Delta
D_x(\omega){\rm d}\mu_\omega$ holds for all $\Delta\in{\mathcal B}$.
Note that, $D_{\alpha x+y}(\omega)=\alpha D_x(\omega)+D_y(\omega)$
a.e. $\mu$. Hence there exists an operator-valued function $D$
defined on $ \Omega $ so that $D(\omega)x=D_x(\omega)$ a.e. $\mu$
for each $x\in H$. And then ${\rm (B)}\int_\Delta D(\omega)x{\rm
d}\mu_\omega={\rm (B)}\int_\Delta D_x(\omega){\rm d}\mu_\omega$ for
any $x\in H$. Since,   $\int_\Delta \langle x|D(\omega)|x\rangle{\rm
d}\mu_\omega =\langle x|E(\Delta)|x\rangle\geq 0$ for any Borel set
$\Delta$, one sees that $\langle x|D(\omega)|x\rangle\geq 0$ a.e.
$\mu$ for each $x\in H$. So, almost all $D(\omega)$ are  (may
unbounded) operators with domain $H$ satisfying $\langle
x|D(\omega)|x\rangle\geq 0$ and $$ E(\Delta)x={\rm (B)}\int_\Delta
D(\omega)x{\rm d}\mu_\omega \eqno(A.2)
$$
holds for all $x\in H$ and $\Delta\in\mathcal B$.\hfill$\Box$

Some times we denote the relation in Eq.(A.2) by $$E(\Delta)={\rm
(SOT)}\int_\Delta D(\omega){\rm d}\mu_\omega \eqno(A.3)$$ holds for
any $\Delta\in\mathcal B$.

\section*{{\bf Appendix B}: Proof of main result}

Now we are at a position to give a proof of the main result theorem
1.

 {\bf Proof of Theorem 1.}\quad Assume that $\rho\in{\mathcal
S}(H_A\otimes H_B)$ is a semi-SSPPT state. We have to show that
$\rho$ is separable. We only need to check the case that $\rho$ is
semi-SSPPT up to part B since the proof for the case of semi-SSPPT
up to A is similar.

As $\rho$ is a semi-SSPPT state up to part B, we may write
$\rho=X^\dag X$, where $X$  upper triangular operator matrices of
the form mentioned in Eq.(4)  with respect to an orthonormal basis
$\{|i_a\rangle\}$ of $H_A$.  Let $C_k$ be the operator matrix with
the same size as that of $X$, which is induced from $X$ by replacing
all entries by zero except for the $k$th row of $X$, i.e.,
\begin{eqnarray*}
C_k=\left(\begin{array}{c|c|c|c|c|c|c|c|c}
0&\cdots&0&0&0&0&\cdots&0&\cdots \\ \hline
\vdots&\ddots&\vdots&\vdots&\vdots
&\vdots&\vdots&0&\cdots \\ \hline
0&\cdots&0&0&0&0&\cdots&0&\cdots \\ \hline
0&\cdots&0&X_k&S_{k,k+1}X_k&S_{k,k+2}X_k
&\cdots&S_{km}X_k&\cdots \\ \hline
0&\cdots&0&0&0&0&\cdots&0&\cdots \\ \hline
\vdots&\cdots&\vdots&\vdots&\vdots&\vdots
&\cdots&\vdots&\ddots\end{array}\right),
\end{eqnarray*}
$ k=1, 2, \dots$. Then $C_k$ is a Hilbert-Schmidt operator and
\begin{eqnarray}
\rho=\sum_{k}C_k^\dag C_k,
\end{eqnarray}
Here the series converges in the trace-norm. If $C_k\not=0, $ write
$C_k^\dag C_k=p_k\rho_k$ where $ p_k={\rm Tr}(C_k^\dag C_k)$. We
will show that $\rho_k$ is separable for any $k$ whenever
$C_k\not=0$, and then, $\rho=\sum_k p_k\rho_k$ is separable, too.

Consider the case when $k=1$. We have
$$
p_1\rho_1=(X_1^\dag S_{1i}^\dag
S_{1j}X_1)=\sum_{i,j}(|i_a\rangle\langle j_a|)\otimes (X_1^\dag
S_{1i}^\dag S_{1j}X_1) \eqno(B.1)
$$
with $\quad S_{11}=I_B$. \if false
$$\begin{array}{lr}p_1\rho_1=&\\
\left(\begin{array}{c|c|c|c|c|c}
X_1^\dag X_1&X_1^\dag S_{12}X_1&X_1^\dag S_{13}X_1
&\cdots&X_1^\dag S_{1m}X_1 &\cdots\\ \hline
X_1^\dag S_{12}^\dag X_1&X_1^\dag S_{12}^\dag S_{12}X_1
&X_1^\dag S_{12}^\dag S_{13}X_1&\cdots
&X_1^\dag S_{12}^\dag S_{1m}X_1 &\cdots\\ \hline
\vdots&\vdots&\vdots&\vdots&\vdots &\vdots\\ \hline
X_1^\dag S_{1,m-1}^\dag X_1&X_1^\dag S_{1,m-1}^\dag S_{12}X_1
&X_1^\dag S_{1,m-1}^\dag S_{13}X_1&\cdots
&X_1^\dag S_{1,m-1}^\dag S_{1m}X_1&\cdots\\ \hline
X_1^\dag S_{1m}^\dag X_1&X_1^\dag S_{1m}^\dag S_{12}X_1
&X_1^\dag S_{1m}^\dag S_{13}X_1&\cdots
&X_1^\dag S_{1m}^\dag S_{1m}X_1&\cdots\\ \hline
\vdots&\vdots&\vdots&\cdots&\vdots&\ddots
\end{array}\right).&
\end{array}$$
~~\fi Since $\rho$ is semi-SSPPT up to part B, $\{S_{1i}\}$ is a
commutative set of normal operators. Then there exists a normal
operator $N_1\in{\mathcal B}(H_B)$ and bounded Borel functions
$\{f_{1i}\}$ such that $S_{1i}=f_{1i}(N_1)$. Let
$N_1=\int_{\sigma(N_1)} \omega {\rm d}E_\omega$ be the spectral
decomposition, where $(\sigma(N_1), {\mathcal B}, H,E)$ is the
spectral measure of $N_1$, $\mathcal B$ is the $\sigma$-algebra of
all Borel subsets of $\sigma(N_1)$, the spectrum of $N_1$. Thus, we
have $S_{1i}=\int_{\sigma(N_1)} f_{1i}(\omega){\rm d} E_\omega$.
Because $H_B$ is separable, there exists a probability measure, that
is, the scalar spectral measure, $(\sigma(N_1), {\mathcal B}, \mu)$
so that, for any $\Delta\in{\mathcal B}$,  $E(\Delta)=0$ if and only
if $\mu(\Delta)=0$ (Ref. \cite{Co}). Then, by Proposition A.3, there
exists an operator-valued function $D$ such that $\langle
x|D(\omega)|x\rangle\geq 0$ a.e. $\mu$ for each $x\in H$ and
\begin{eqnarray*}
E(\Delta)={\rm (SOT)}\int_\Delta D(\omega) {\rm d}\mu_\omega
\end{eqnarray*}
holds for all $\Delta\in{\mathcal B}$. By Eq.(B.1) one gets, for
each product vector $x_A\otimes x_B\in H_A\otimes H_B$,
$$\begin{array}{rl} & \rho_1(x_A\otimes x_B) \\=& p_1^{-1}
\sum_{i,j}(|i_a\rangle\langle j_a|)\otimes (X_1^\dag
\int_{\sigma(N_1)} f_{1i}(\omega)^*
f_{1j}(\omega){\rm d}E_\omega X_1)(x_A\otimes x_B)\\
=& p_1^{-1} \sum_{i,j}(|i_a\rangle\langle j_a|)x_A\otimes ( {\rm
(B)}\int_{\sigma(N_1)} f_{1i}(\omega)^*
f_{1j}(\omega)X_1^\dag D(\omega)X_1x_B{\rm d}\mu_\omega)\\
=& \sum_{i,j}\{{\rm (B)}\int_{\sigma(N_1)} [
p_1^{-1}f_{1i}(\omega)^*f_{1j}(\omega)|i_a\rangle\langle
j_a\rangle]x_A\otimes [X_1^\dag D(\omega)X_1]x_B{\rm d}\mu_\omega\}.
\end{array}
$$
Take an orthonormal basis $\{|j_b\rangle\}$ of $H_B$. For any $n$,
let $P_n$ be the $n$-rank projection onto
span$\{|i_a\rangle\}_{i=1}^n$ and $Q_n$ be the $n$-rank projection
onto span$\{|j_b\rangle\}_{j=1}^n$. Then
$$\begin{array}{rl}
& (P_n\otimes Q_n)\rho_1(P_n\otimes Q_n)(x_A\otimes x_B)\\
= &{\rm (B)}\int_{\sigma(N_1)} [ p_1^{-1}
\sum_{i,j=1}^nf_{1i}(\omega)^*f_{1j}(\omega)|i_a\rangle\langle
j_a\rangle]\otimes [Q_nX_1^\dag D(\omega)X_1Q_n]{\rm
d}\mu_\omega(x_A\otimes x_B)
\end{array}
$$
holds for any $x_A\otimes x_B\in H_A\otimes H_B$, which entails that
$$\begin{array}{rl}
&(P_n\otimes Q_n)\rho_1(P_n\otimes Q_n)\\ =& {\rm
(B)}\int_{\sigma(N_1)} [ p_1^{-1}
\sum_{i,j=1}^nf_{1i}(\omega)^*f_{1j}(\omega)|i_a\rangle\langle
j_a\rangle]\otimes [Q_nX_1^\dag D(\omega)X_1Q_n]{\rm
d}\mu_\omega.\end{array}
$$
Let $A_n(\omega)=\sum_{i,j=1}^n
p_1^{-1}f_{1i}(\omega)^*f_{1j}(\omega)|i_a\rangle\langle
j_a\rangle]$ and $B_n(\omega)=Q_nX_1^\dag D(\omega)X_1Q_n$. It is
easily seen that $A_n(\omega)$ is a rank one positive operator on
$H_A$ for each $\omega\in\sigma(N_1)$. As $B_n(\omega)$ is bounded
and satisfies $\langle x_B|B_n(\omega)|x_B\rangle \geq 0$ for any
$x_B\in H_n$,  $B_n(\omega)$ must be a positive operator. Therefore,
$$\begin{array}{rl}\sigma_n=&\frac{1}{{\rm Tr}((P_n\otimes Q_n)\rho_1(P_n\otimes Q_n))}(P_n\otimes Q_n)\rho_1(P_n\otimes
Q_n)\\
=& \frac{1}{{\rm Tr}((P_n\otimes Q_n)\rho_1(P_n\otimes Q_n))}{\rm
(B)}\int_{\sigma(N_1)}A_n(\omega)\otimes B_n(\omega)\ {\rm
d}\mu_\omega\end{array}$$ is a separable state. Since
$\rho_1=\mbox{\rm SOT-}\lim_{n\to\infty}(P_n\otimes
Q_n)\rho_1(P_n\otimes Q_n)$, we see that $\rho_1$ is SOT-separable.
Then the proposition A.2 ensures that $\rho_1$ is a separable state,
as desired.

Similarly, one can check that $\rho_k$ is   separable for each $k$,
$k\geq1$. Hence, we see that $\rho$ is a  separable state, as
desired.\hfill$\Box$

\end{document}